\documentclass[a4paper,11pt]{article}
\usepackage{pos}

\usepackage{amsmath}
\usepackage[separate-uncertainty=true,per-mode=symbol]{siunitx}
\usepackage{setspace}

\newcommand{\refeq}[1]{Eq.~(\ref{#1})}

\newcommand{\reffig}[1]{Fig.~\ref{#1}}

\newcommand{\refsec}[1]{Section~\ref{#1}}

\newcommand{\reftab}[1]{Table~\ref{#1}}
\newcommand{\refref}[1]{Ref.~\cite{#1}}

\title{Parameterization of Muon Production Profiles in the Atmosphere}
 \ShortTitle{Parameterization of muon production profiles}

\author*[a]{Stef Verpoest}
\author[b]{Thomas K. Gaisser}

\affiliation[a]{Dept. of Physics and Astronomy, University of Gent, B-9000 Gent, Belgium}

\affiliation[b]{Bartol Research Institute and Dept. of Physics and Astronomy, University of Delaware, Newark, DE 19716, USA}

\emailAdd{stef.verpoest@ugent.be}
\emailAdd{gaisser@udel.edu}

\abstract{Production of high-energy muons in cosmic-ray air showers, relevant for underground detectors, depends on the properties of the primary cosmic ray as well as the atmospheric temperature through the competition between decay and re-interaction of charged pions and kaons. We present a parameterization of muon production profiles based on simulations as a function of the primary cosmic-ray energy, mass and zenith angle, the minimum energy for a muon to reach the detector and an atmospheric temperature profile. We illustrate how this can be used to calculate muon bundle properties such as multiplicity and transverse size and their seasonal variations in the context of underground measurements in coincidence with a surface detector which fixes the primary cosmic-ray energy.}

\FullConference{37$^{\rm{th}}$ International Cosmic Ray Conference (ICRC 2021)\\
		July 12th -- 23rd, 2021\\
		Online -- Berlin, Germany}

\begin{document}
\maketitle

\section{Introduction}\label{sec:intro}

The yield of high-energy muons in air showers induced by cosmic rays interacting near the top of the atmosphere is relevant for understanding event rates and properties of muon bundles in underground detectors. The following formula, originally proposed by Elbert~\cite{Elbert:1979abc}, has been used to estimate the average multiplicity $\langle N_\mu \rangle$ of muons above a certain energy $E_\mu$:
\begin{equation}
\langle N_\mu(>E_\mu, E_0, A, \theta)\rangle\;\approx\;A\times\frac{K}{ E_\mu\,\cos\theta}\,\left (
\frac{E_0}{ A\,E_\mu}\right )^{\alpha_1}\,
\left (1\,-\,\frac{A\,E_\mu}{E_0}\right)^{\alpha_2},
\label{eq:ElbertFormula}
\end{equation}
where $E_0$, $A$, and $\theta$ are respectively the energy, mass number, and zenith of the primary cosmic ray nucleus. The normalization constant $K$ and exponents $\alpha_1$ and $\alpha_2$ are to be derived from simulations. The scaling with $AE_\mu / E_0$ follows from the superposition approximation, which assumes that an incident nucleus of mass $A$ and energy $E_0$ can be treated as $A$ independent nucleons of energy $E_0/A$.

The Elbert formula does not describe the fact that muon production depends on the density (or temperature) of the atmosphere through the competition between re-interaction and decay of the parent mesons. In summer, when the atmosphere is warmer and less dense, more mesons will decay to muons rather than interact, and the number of high-energy muons in the shower will be larger. In this work, we present a generalization of the Elbert formula describing the production of muons above some energy threshold as a function of slant depth in the atmosphere, based on a parameterization of simulations, and including factors taking into account the atmospheric temperature (\refsec{sec:param}). This parameterization allows one to estimate not only the multiplicity of muon bundles in air showers but also its transverse size and the seasonal variations of both these properties, which we illustrate for the case of IceCube~\cite{Aartsen:2016nxy} in \refsec{sec:seasonal}. 

Other applications include the calculation of event rates of single- and multiple-muon events in underground detectors, where one integrates over the spectrum of primary nucleons~\cite{gaisser2021profiles}, but are not discussed here.

\section{Muon production profiles}\label{sec:param}

The production of muons above a certain energy threshold, differential in slant depth throughout the atmosphere along the shower axis, is referred to as the longitudinal production profile. The idea is to perform a large number of air-shower simulations and to obtain the average muon production profile for primary cosmic rays with energy $E_0$, mass number $A$, zenith angle $\theta$, and for muons with energy above $E_\mu$. We have used CORSIKA v7.7100~\cite{CORSIKA_Heck} using Sibyll 2.3c~\cite{Riehn:2019jet} as the high-energy interaction model, and an atmospheric profile describing the average South Pole atmosphere in April between 2007 and 2011~\cite{DeRidder:2019}. To the average profiles obtained from simulation, we fit a function of the following form, which we explain below,
\begin{align}
\begin{split}
&\left\langle\frac{\mathrm{d}N}{\mathrm{d}X}(>E_\mu, X, T, E_0, A, \theta)\right\rangle =\\
&\qquad\qquad N_{max} \times \exp \left((X_{max} - X)/\lambda\right) \times \left(\frac{X_0 - X}{X_0 - X_{max}}\right)^{(X_{max}-X_0)/\lambda} \times \frac{X_{max}-X}{ \lambda(X-X_0)} \\
&\qquad\qquad\times \left[ 0.92 \times \frac{r_\pi \lambda_\pi \epsilon_\pi}{fE_\mu \cos(\theta) X} \times \frac{1}{1+\frac{r_\pi \lambda_\pi \epsilon_\pi}{fE_\mu \cos(\theta) X}} + 0.08 \times \frac{r_K \lambda_K \epsilon_K}{fE_\mu \cos(\theta) X} \times \frac{1}{1+\frac{r_K \lambda_K \epsilon_K}{fE_\mu \cos(\theta) X}}\right] \\
&\qquad\qquad \times \left(1 - \frac{A E_\mu}{E_0}\right)^{5.99},
\end{split}
\label{eq:formula}
\end{align}
where $T$ is the temperature at a slant depth $X$. 

The first line on the right-hand side is the derivative of the Gaisser-Hillas (G-H) function~\cite{GaisserHillasabc}, which we interpret as the rate of production of charged mesons per d$X$ (\si{\g \per \cm \squared}). The parameters of number of particles at shower maximum ($N_\mathrm{max}$), depth of shower maximum ($X_\mathrm{max}$), depth of first interaction ($X_0$), and interaction length ($\lambda$) are the free parameters during the fit and, as they are applied here to the charged mesons in the hadronic cascade, their numerical values are quite different from those of the original G-H function. 

\begin{figure}
    \centering
    \includegraphics[width=0.5\textwidth]{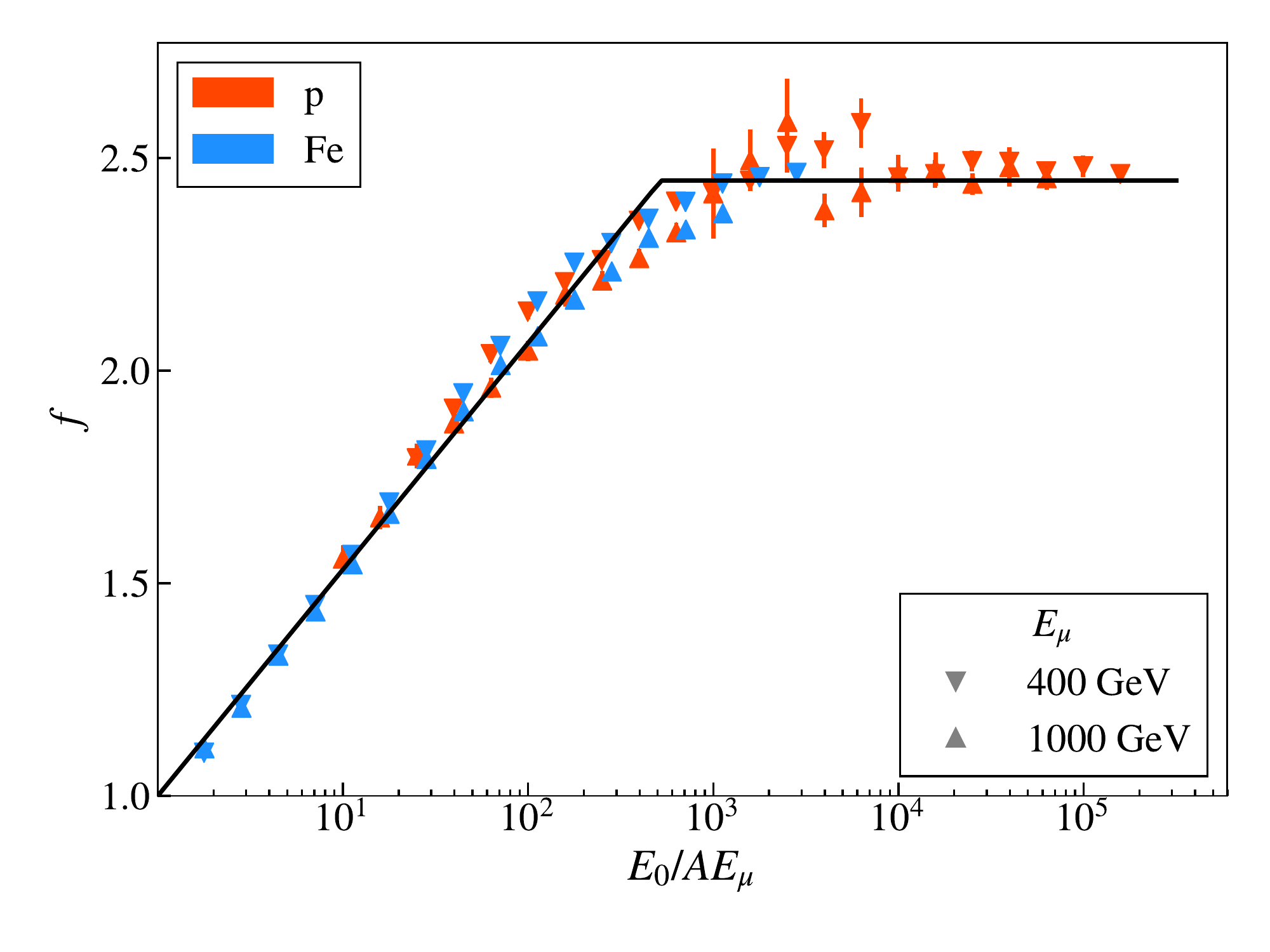}
    \caption{Ratio between the mean energy of muons above the threshold and the muon threshold energy $E_\mu$. Markers are derived from vertical proton and iron shower simulations. Our approximation of $f$ used in \refeq{eq:formula} is given by the black line.}
    \label{fig:factor}
\end{figure}

In the second line of \refeq{eq:formula}, we multiply by the probability for mesons to decay to a muon relative to the total rate of decay and re-interaction. We consider two channels for muon production, namely decay of charged pions and kaons $\pi^\pm/ K^\pm \rightarrow \mu + \nu_\mu$, with branching ratios of 100\% and 63.5\% respectively. The decay fraction for charged pions with interaction length $\lambda_\pi$ and decay length $d_\pi$ is 
\begin{equation}
    \frac{1/d_\pi}{1/d_\pi + 1/\lambda_\pi}.
\end{equation}
The decay length is given by~\cite{Gaisser:2016uoy}
\begin{equation}
    \frac{1}{d_\pi} = \frac{\epsilon_\pi}{E_\pi \cos \theta X},
\end{equation}
where $E_\pi$ is the energy of the pion and $\epsilon_\pi$ the pion critical energy given by
\begin{equation}
    \epsilon_\pi = \frac{m_\pi c^2}{c \tau_\pi} \frac{RT}{Mg} \approx \SI{115}{\GeV} \times \frac{T}{\SI{220}{\K}},
\end{equation}
with $c$ the speed of light in vacuum, $m_\pi$ and $\tau_\pi$ the pion mass and lifetime, R the molar gas constant, M the molar mass of the atmosphere, and $g$ the gravitional constant. On average, the muon that results from pion decay has an energy $E_\mu = r_\pi \times E_\pi$ with $r_\pi \approx 0.79$. For kaons, the critical energy is larger by a factor of 7.45 because of its larger mass and shorter decay length, and the muon energy in this case is defined by $r_K \approx 0.52$. The factors of 0.92 and 0.08 preceding the pion and kaon terms are the relative fractions of momentum carried by charged pions and kaons after taking into account the branching ratios. The momentum fraction carried by charged pions and kaons in p-air interactions is given by Fig.~5.2 of Ref.~\cite{Gaisser:2016uoy} as 0.29 and 0.040 respectively. Combined with the branching ratios, this gives $0.29 / (1\times0.29 + 0.635\times0.04) = 0.92$ for charged pions and 0.08 for charged kaons. To take into account the fact that the mean energy of muons is larger than the threshold muon energy itself, we replace $E_\mu$ by $f E_\mu$, where the factor $f$ gives the ratio between the mean energy of muons above the threshold energy and the threshold energy $E_\mu$. Its behaviour can be derived from simulations and is shown in \reffig{fig:factor} for the muon energy range we consider. It has a piecewise behaviour parametrized by the black line, with the parameters included in \reftab{tab:params}.

The third line of \refeq{eq:formula} is the threshold factor from the Elbert formula \refeq{eq:ElbertFormula}, with an exponent fitted to our simulations. It describes the suppression in muon multiplicity when the energy per nucleon is close to the minimum muon energy.

\begin{figure}
    \centering
    \includegraphics[width=0.5\textwidth]{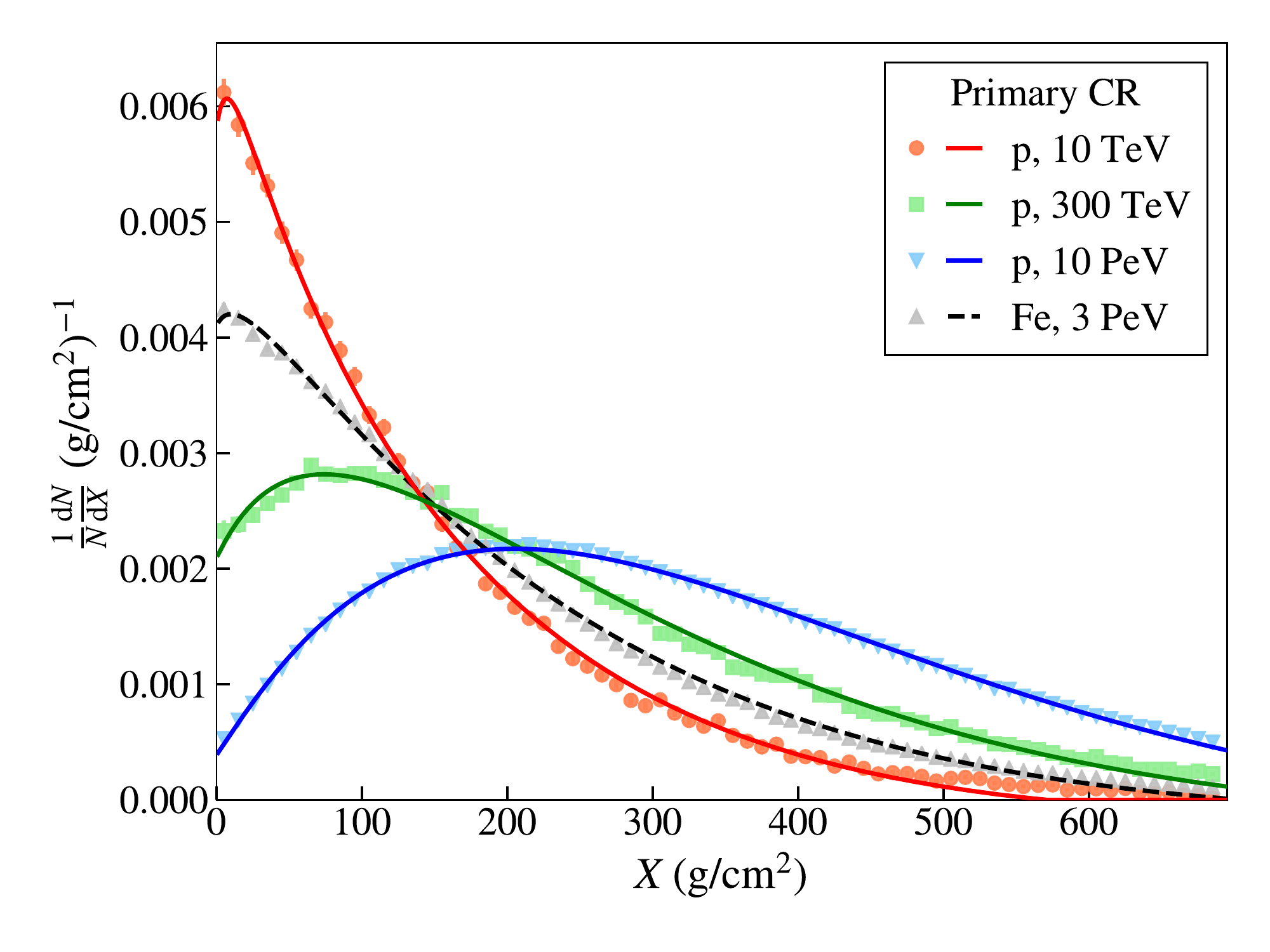}
    \caption{Normalized muon production profiles for vertical showers with $E_\mu > \SI{300}{\GeV}$. The markers show the average profiles obtained from simulations, the lines are the fits of \refeq{eq:formula} to the simulation results.}
    \label{fig:profiles}
\end{figure}

\begin{figure}
    \centering
    \includegraphics[width=.9\textwidth]{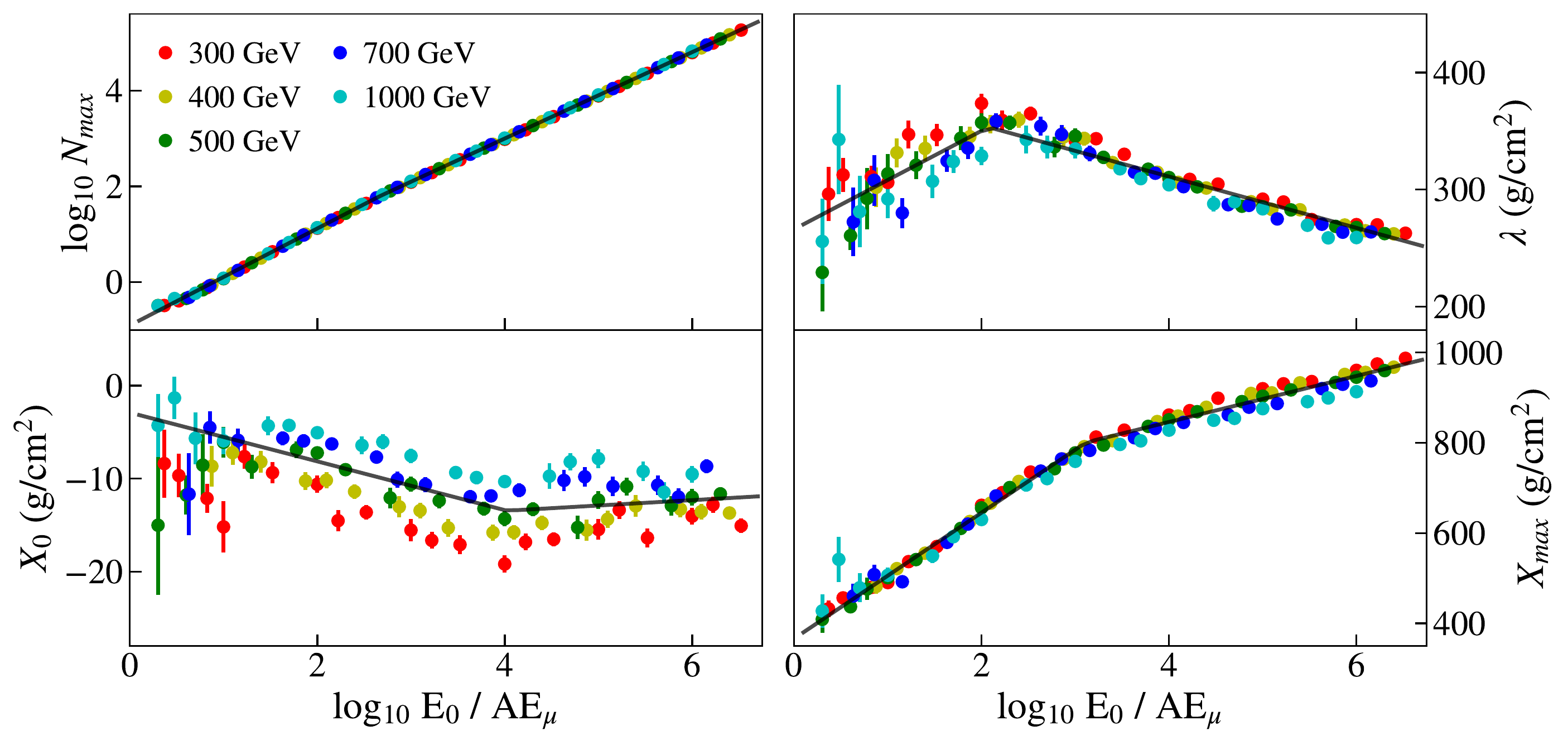}
    \caption{Optimal values of the fit parameters $N_{max}$, $X_{max}$, $\lambda$, and $X_0$ of Eq.~\ref{eq:formula}, as obtained from fits to vertical proton showers for various minimum muon energies $E_\mu$ over a large range of primary energies. The black lines are fits to these results with the functions of \refeq{eq:params}, resulting in the parameters given in \reftab{tab:params}.}
    \label{fig:params}
\end{figure}

Examples of the formula of \refeq{eq:formula} fit to production profiles derived from CORSIKA simulations are shown in \reffig{fig:profiles} for $E_\mu > \SI{300}{\GeV}$. We have repeated this procedure for muon threshold energies of 300, 400, 500, 700, and \SI{1000}{\GeV} and a large range of primary energies. The optimized values of $N_{\mathrm{max}}$, $X_{\mathrm{max}}$, $\lambda$, and $X_0$ for vertical proton showers are shown in \reffig{fig:params}. We observe that their behaviour depends in leading order on $E_0/AE_\mu$, and parametrize it with the following functions,
\begin{align}
\begin{split}
\label{eq:params}
N_{\rm max}&= c_i\times A \times \left(\frac{E_0}{AE_\mu}\right)^{p_i}\\
X_{\rm max}, \lambda, X_0&=a_i + b_i\times \log_{10}\left(\frac{E_0}{AE_\mu}\right),
\end{split}
\end{align}
where $c_i$, $p_i$, $a_i$, and $b_i$ are defined for each function separately and have two regimes with a break
at $R_b = \frac{E_0}{A E_\mu} = 10^{q}$ and parameters $(a_i,b_i)$ with $i=1$
below the break and $i=2$ above. The resulting parameters are listed in \reftab{tab:params}. A simple Python implementation of this parameterization is made available on Github\footnote{\url{https://github.com/verpoest/muon-profile-parameterization}}. Note that the scaling with $E_0/AE_\mu$ is not perfect; a remaining dependence on $E_\mu$ can be observed in \reffig{fig:params}. It is therefore recommended to optimize the simulations and fits to the energy regime relevant for the application or detector that is studied.

\begin{table}[htb]
\begin{center}
\caption{Parameter values for \refeq{eq:params} for $\SI{300}{\GeV} \lesssim E_\mu \lesssim \SI{1}{\tera \eV}$.}
\begin{tabular}{lr|r|r|r}
\hline \hline
&$i$& $c_i$ & $p_i$ & $q$\\
\hline \hline
$N_{\rm max}$ &1&0.124&1.012&2.677 \\
&2&0.244&0.902& \\
\hline \hline
& $i$ & $a_i$ (g/cm$^2$)& $b_i$ (g/cm$^2$)& $q$ \\
\hline \hline
$X_{\rm max}$ &1& 366.2 & 139.5 & 3.117 \\
&2& 642.2 & 51.0 & \\
\hline
$\lambda$ &1& 266.0 & 42.1 & 2.074 \\
&2& 398.8 & -21.9 & \\
\hline
$X_0$ & 1& -2.9 & -2.6 & 4.025 \\
&2& -15.8 & 0.6 & \\
\hline
$f$ & 1 & 1 & 0.53 & 2.72 \\
& 2 & 2.45 & - \\
\hline \hline
\end{tabular}
\label{tab:params}
\end{center}
\end{table}

\section{Seasonal variations of muon bundle properties}\label{sec:seasonal}

As an example of the application of the parameterization of \refsec{sec:param}, we will examine the case where muon bundles are observed in an underground detector and the primary cosmic ray energy is determined independently by a surface detector. We will use values relevant for air showers detected coincident between the surface array IceTop~\cite{IceCube:2012nn}, located at the South Pole, which detects air showers with primary energies between \SI{1}{\peta \eV} and \SI{1}{\exa \eV} at an atmospheric depth of roughly \SI{700}{\g \per \cm \squared}, and IceCube\cite{Aartsen:2016nxy}, which sits vertically below IceTop buried under \SI{1.5}{\km} of ice and allows for the detection of muons above approximately \SI{400}{\GeV}. The calculations are performed using atmospheric data for the South Pole obtained from the AIRS satellite~\cite{NASA:2018abc}, which provides the temperature at different, unevenly spaced, atmospheric pressure levels between \SI{1}{\hecto \Pa} and \SI{700}{\hecto \Pa}. These pressures are converted to atmospheric depth and interpolated to a regular grid.

\begin{figure}
    \centering
    \includegraphics[width=0.5\textwidth]{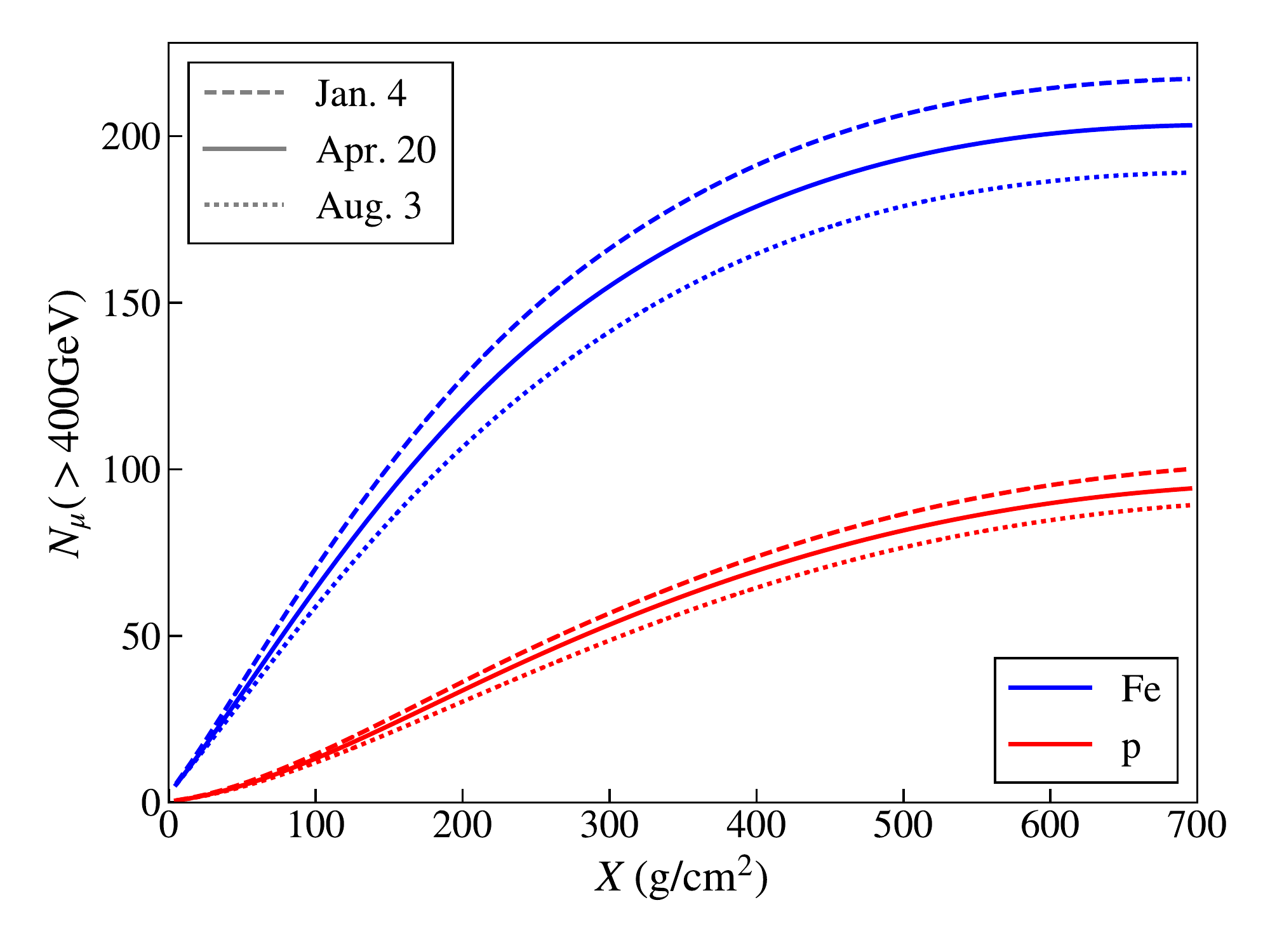}\includegraphics[width=0.5\textwidth]{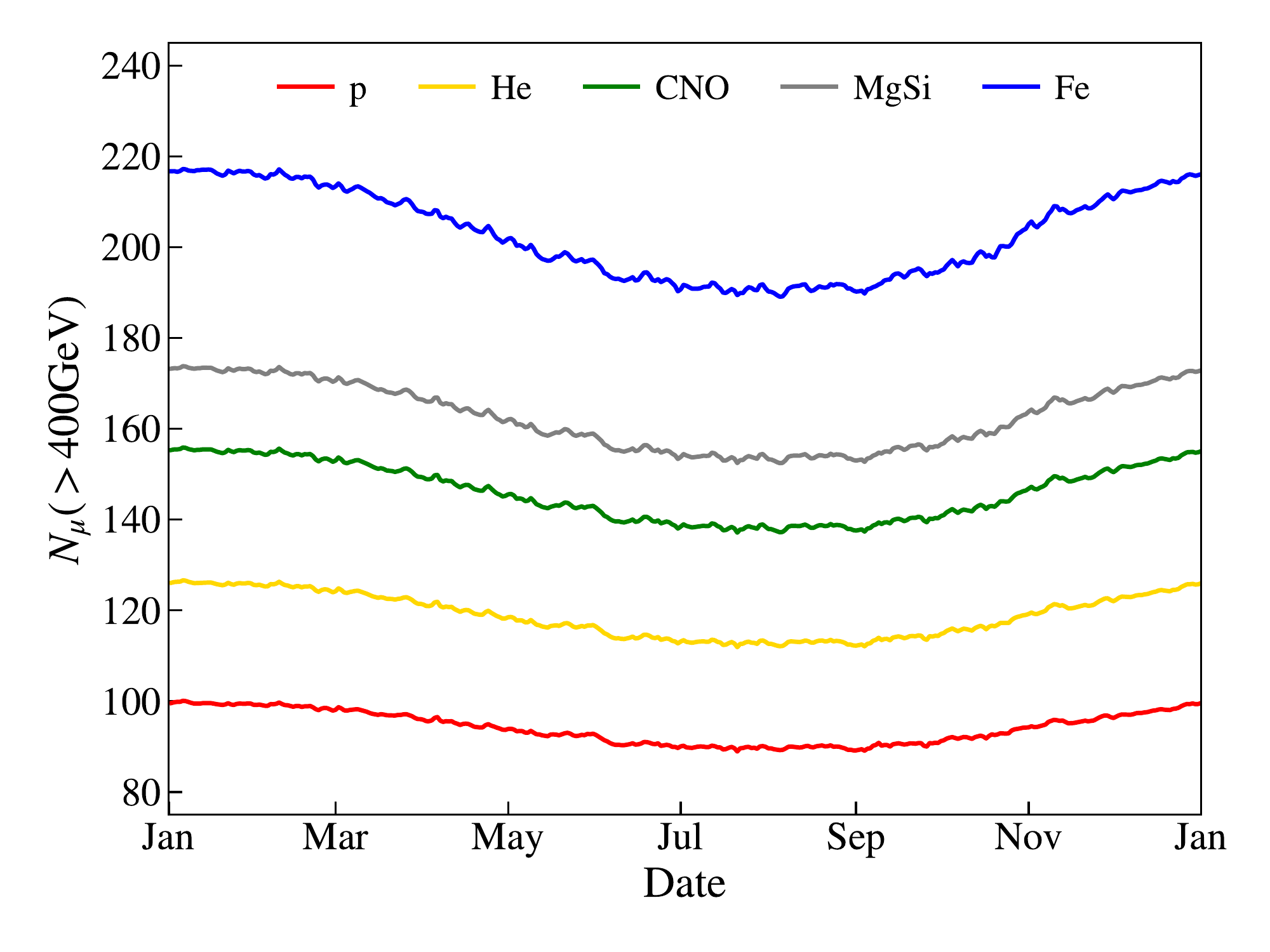}
    \caption{Left: Integral production profiles for \SI{10}{\peta\eV} proton and iron showers for three days in 2017 at the South Pole where the expected muon multiplicity is approximately minimal, maximal and average. Right: Variation of the expected multiplicity throughout the year for vertical \SI{10}{\peta\eV} showers from five primary mass groups.}
    \label{fig:mult_var}
\end{figure}

Using the temperature profiles together with the parameterization, we obtain muon production profiles which can be integrated to find the expected muon multiplicity. \reffig{fig:mult_var} shows the expected multiplicity of muons above \SI{400}{\GeV} in \SI{10}{\peta \eV} vertical showers throughout the year 2017, as well as the integral profiles for three days corresponding roughly to the days with minimal, average, and maximal multiplicity. It can be seen that the multiplicity is maximal in the austral summer, when temperatures are highest. The calculation predicts a seasonal variation of about 6\% around the mean. This may be an important uncertainty to consider in cosmic-ray composition analyses based on muon bundle measurements~\cite{IceCube:2019hmk}.

\begin{figure}
    \centering
    \includegraphics[width=0.5\textwidth]{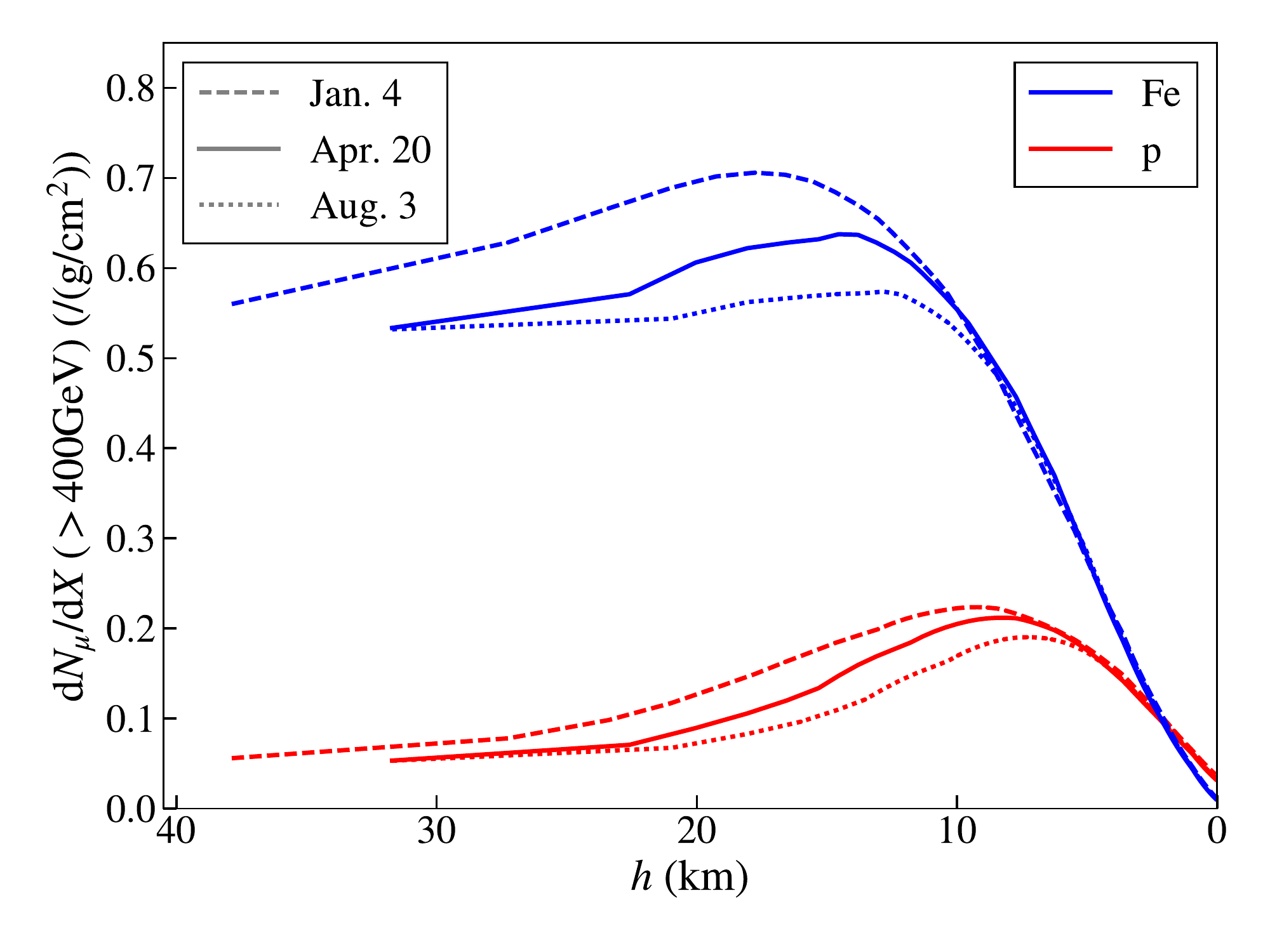}\includegraphics[width=0.5\textwidth]{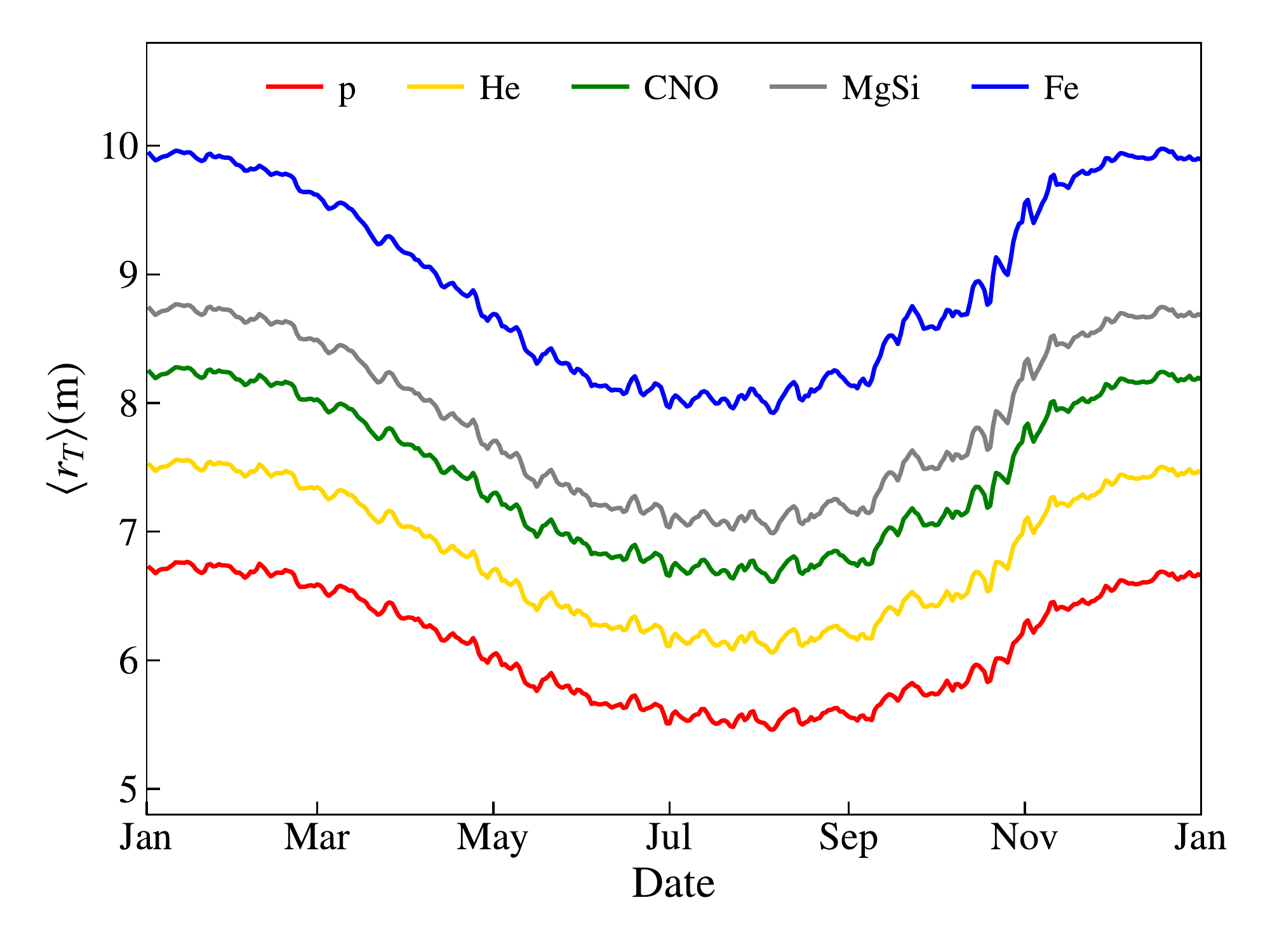}
    \caption{Left: Differential muon production versus altitude for three different days at the South Pole, measured relative to the surface above IceCube, in vertical \SI{10}{\peta\eV} showers. Right: Seasonal variation of the estimated transverse size of the muon bundle (altitude effect only) for five mass groups.}
    \label{fig:size}
\end{figure}

Because the parameterization describes the muon production as function of slant depth in the atmosphere, it is also possible to extract information about the altitude of production and to estimate the transverse size of a muon bundle. A muon with energy $E_\mu$ produced at an altitude $h$ with a transverse momentum $p_T$ will have a transverse distance from the shower axis given by

\begin{equation}
    r_T = \frac{p_T}{E_\mu} \times \frac{h}{\cos \theta},
    \label{eq:transverse_dist}
\end{equation}
where $\theta$ is the zenith angle of the primary. At a vertical depth $X_v$ the atmospheric pressure is $P = gX_v$ and the density is given by $\rho = -\mathrm{d}X_v/\mathrm{d}h$. Assuming the ideal gas law, one can calculate the altitude corresponding to vertical depth $X_v$ as
\begin{equation}
    h(X_v) = \frac{RT}{Mg} \ln \frac{X_0}{X_v},
\end{equation}
where $X_0$ is the vertical depth at $h=0$. Using this, we will perform a simple estimate of the expected bundle size, assuming a mean value of transverse momentum for the muons of $\langle p_T \rangle \approx \SI{350}{\mega \eV}$~\cite{Alper:1974xp}. As zero-point $h=0$ for the altitude we use the surface above the IceCube detector, located at an elevation of \SI{2835}{\m} with an atmospheric depth $X_0 \approx \SI{700}{\g \per \cm \squared}$. The left panel of \reffig{fig:size} shows the differential muon production as function of altitude for vertical proton and iron showers at three different days corresponding again roughly to the yearly average and two extremal days. It is clear that muons are produced higher in the atmosphere for heavier nuclei. For a given primary mass, production happens at higher altitude in the summer compared to colder days because of the thermal expansion of the atmosphere. An estimate of the expected bundle size $\langle r_T \rangle$ is obtained by taking the weighted average of the transverse distance for a muon with $\langle p_T \rangle$ at a depth $X$ using \refeq{eq:transverse_dist}, multiplying it with the production profile and integrating over depth. The result is shown in the right panel of \reffig{fig:size}, where we see that the muon bundle has the largest spread in the warmest months, corresponding to the higher production altitude. The magnitude of the seasonal variations is roughly 10\% around the average value.

Note that we report the multiplicity and transverse size of the muon bundle at the surface above the IceCube detector. The estimate of the transverse size is also limited to the geometrical effect. A full estimate of the muon bundle properties in the detector needs to take into account propagation through the overburden, where multiple scattering of the muons will further increase the spread of the muons~\cite{Lipari:1991ut}. Also separation of muons by bending in the geomagnetic field before they reach the surface can be important, especially for inclined showers~\cite{Abreu:2011ki}.

\section{Summary}\label{sec:summary}

We have presented a parameterization of muon production profiles in cosmic-ray air showers based on fits to air-shower simulations. The production profile for a certain primary cosmic ray and a muon energy threshold can be obtained for realistic atmospheres to estimate the muon multiplicity and the transverse size of the muon bundle caused by the geometrical separation related to the muon production altitude. Because the temperature dependence of the decay probability of parent mesons is included in \refeq{eq:formula}, the seasonal variations of these quantities can be determined. An estimate performed at fixed primary energy relevant for the case of IceCube shows that the multiplicity and the transverse size are maximal when the atmosphere is at its warmest, consistent with the increased decay rate and higher muon production altitude resulting from the thermal expansion of the atmosphere, and vice-versa when the atmosphere is colder. Because the parameterization does not scale perfectly with the ratio of the muon energy and primary nucleon energy, it should be optimized for detectors with different conditions, e.g. the overburden.

Further applications of the parameterization exist but are not included here. One example is the calculation of rates of events of single and multiple muons in underground detectors, as discussed in \refref{gaisser2021profiles}.

\bibliographystyle{ICRC}
\bibliography{references}

\providecommand{\href}[2]{#2}\begingroup\raggedright\setstretch{0.01}\begin{thebibliography}{10}

\bibitem{Elbert:1979abc}
J.~W. Elbert, ``{Multiple muons produced by cosmic ray interactions.},'' in
  {\em {Proceedings of the DUMAND Summer Workshop}}, pp.~101--121.
\newblock {Scripps Institution of Oceanography, La Jolla CA}, 1979.

\bibitem{Aartsen:2016nxy}
{\bfseries IceCube} Collaboration, M.~G. Aartsen {\em et~al.}
  \href{http://dx.doi.org/10.1088/1748-0221/12/03/P03012}{{\em JINST}
  {\bfseries 12} no.~03, (2017) P03012}.

\bibitem{gaisser2021profiles}
T.~K. Gaisser and S.~Verpoest.
\newblock \href{http://arxiv.org/abs/2106.12247}{{\ttfamily arXiv:2106.12247}}.

\bibitem{CORSIKA_Heck}
D.~Heck {\em et~al.}, {\em CORSIKA: A Monte Carlo code to simulate extensive
  air showers, Report FZKA 6019, Forschungszentrum Karlsruhe}, 1998.

\bibitem{Riehn:2019jet}
F.~Riehn, R.~Engel, A.~Fedynitch, T.~K. Gaisser, and T.~Stanev
  \href{http://dx.doi.org/10.1103/PhysRevD.102.063002}{{\em Phys. Rev. D}
  {\bfseries 102} no.~6, (2020) 063002}.

\bibitem{DeRidder:2019}
{De Ridder, Sam}, {\em {Sensitivity of IceCube cosmic ray measurements to the
  hadronic interaction models}}.
\newblock PhD thesis, {Ghent University}, {2019}.

\bibitem{GaisserHillasabc}
T.~Gaisser and A.~Hillas {\em {Proceedings, 15th International Cosmic Ray
  Conference (ICRC1977)}: {Plovdiv, Bulgaria}} {\bfseries 8} (1977) 353--357.

\bibitem{Gaisser:2016uoy}
T.~K. Gaisser, R.~Engel, and E.~Resconi, {\em {Cosmic Rays and Particle
  Physics}}.
\newblock Cambridge University Press,
2016.
\newblock

\bibitem{IceCube:2012nn}
{\bfseries IceCube} Collaboration, R.~Abbasi {\em et~al.}
  \href{http://dx.doi.org/10.1016/j.nima.2012.10.067}{{\em Nucl. Instrum. Meth.
  A} {\bfseries 700} (2013) 188--220}.

\bibitem{NASA:2018abc}
NASA-AIRS.
  \href{http://dx.doi.org/{https://airs.jpl.nasa.gov/data/get-data}}{{\em
  {https://airs.jpl.nasa.gov/data/get-data}} }.

\bibitem{IceCube:2019hmk}
{\bfseries IceCube} Collaboration, M.~G. Aartsen {\em et~al.}
  \href{http://dx.doi.org/10.1103/PhysRevD.100.082002}{{\em Phys. Rev. D}
  {\bfseries 100} no.~8, (2019) 082002}.

\bibitem{Alper:1974xp}
B.~Alper {\em et~al.}
  \href{http://dx.doi.org/10.1016/0370-2693(73)90730-2}{{\em Phys. Lett. B}
  {\bfseries 47} (1973) 275--280}.

\bibitem{Lipari:1991ut}
P.~Lipari and T.~Stanev \href{http://dx.doi.org/10.1103/PhysRevD.44.3543}{{\em
  Phys. Rev. D} {\bfseries 44} (1991) 3543--3554}.

\bibitem{Abreu:2011ki}
{\bfseries Pierre Auger} Collaboration, P.~Abreu {\em et~al.}
  \href{http://dx.doi.org/10.1088/1475-7516/2011/11/022}{{\em JCAP} {\bfseries
  11} (2011) 022}.

\end{thebibliography}\endgroup

\end{document}